\documentclass{article}
\usepackage{natbib}
\usepackage{graphicx}
\usepackage{tabularx}				
\usepackage[euler]{textgreek}		
\usepackage{scalerel,amssymb}
\usepackage[utf8]{inputenc}
\usepackage{booktabs}
\usepackage{amsmath}
\usepackage{multirow}
\usepackage{placeins}
\usepackage[toc,page]{appendix}
\AtBeginEnvironment{appendices}{\crefalias{section}{appendix}}
\usepackage{siunitx}
\usepackage[version=4]{mhchem}
\usepackage{hyperref}
\usepackage[noabbrev,capitalise]{cleveref}
\usepackage{wrapfig}

\makeatletter
\newcommand*{\addFileDependency}[1]{
  \typeout{(#1)}
  \@addtofilelist{#1}
  \IfFileExists{#1}{}{\typeout{No file #1.}}
}
\makeatother

\begin{document}



\title{Machine-learning-optimized perovskite nanoplatelet synthesis}

\author{
  Carola Lampe$^1$\\
  \and
 Ioannis Kouroudis$^2$\\
  \and
    Milan Harth$^2$\\
  \and
 Stefan Martin$^1$\\
  \and  
  Alessio Gagliardi$^2$ \\
  \and
 Alexander S. Urban$^1$ \\
}

\date{%
    $^1$ Nanospectroscopy Group and Center for NanoScience, Nano-Institute Munich, Faculty of Physics \\%
    $^2$ Department of Electrical and Computer Engineering, Technical University of Munich\\[2ex]%
}


\maketitle




\vspace{10mm}









\vspace{10mm}



\begin{abstract}
With the demand for renewable energy and efficient devices rapidly increasing, a need arises to find and optimize novel (nano)materials. This can be an extremely tedious process, often relying significantly on trial and error. Machine learning has emerged recently as a powerful alternative; however, most approaches require a substantial amount of data points, i.e., syntheses. Here, we merge three machine-learning models with Bayesian Optimization and are able to dramatically improve the quality of \ce{CsPbBr3} nanoplatelets (NPLs) using only approximately 200 total syntheses. The algorithm can predict the resulting PL emission maxima of the NPL dispersions based on the precursor ratios, which lead to previously unobtainable 7 and 8 ML NPLs. Aided by heuristic knowledge, the algorithm should be easily applicable to other nanocrystal syntheses and significantly help to identify interesting compositions and rapidly improve their quality. 
\end{abstract}


\section{Introduction}
\begin{figure}[ht]
    \centering
    \includegraphics[width = 0.7\textwidth]{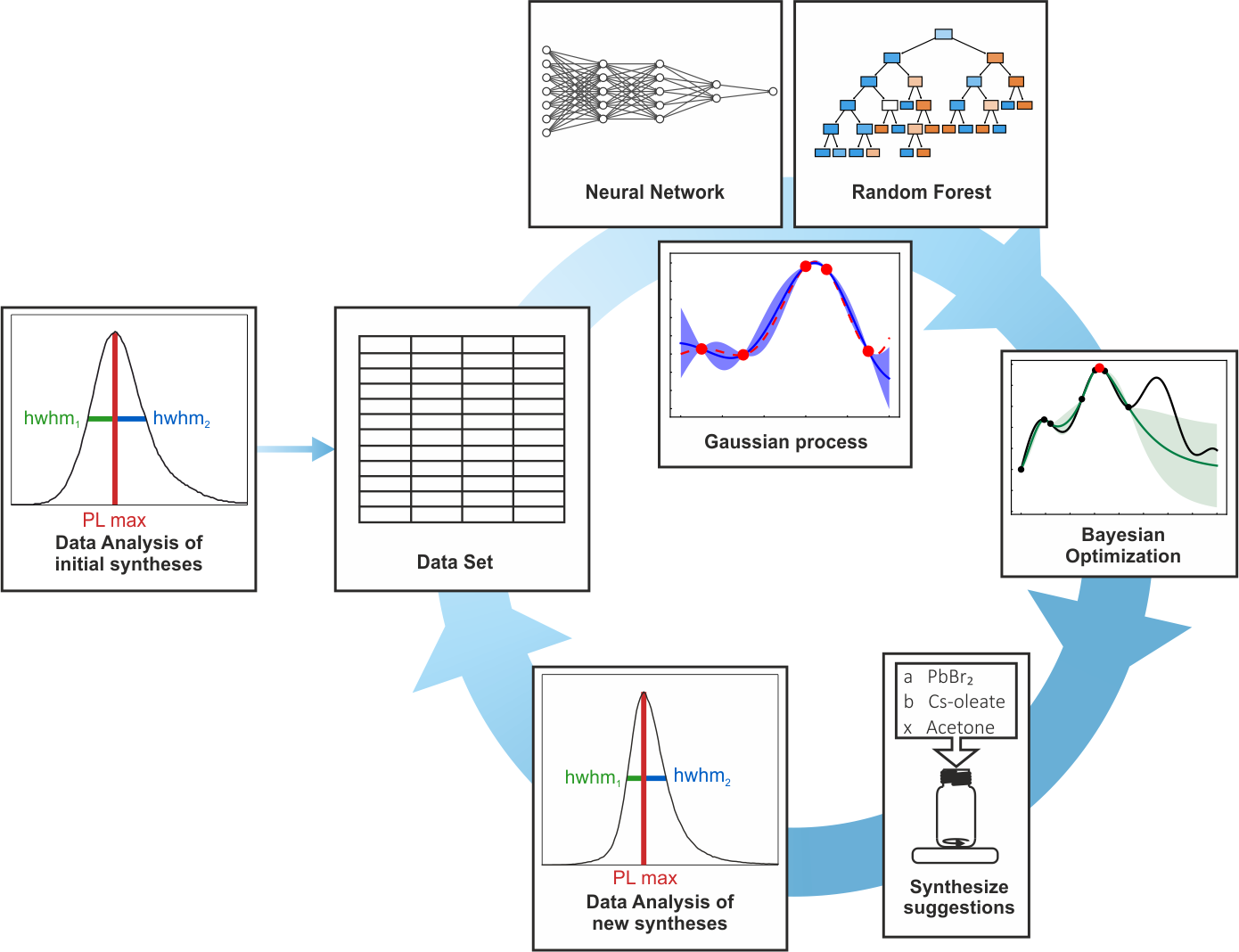}
        \caption{Scheme of the optimization process: Initially existing data points (syntheses) were analyzed and used to predict a spectral figure of merit (FoM) based on the narrowness and symmetry of their PL spectra using Gaussian Processes in combination with a Random Forest and a Neural Network. Constrained Bayesian optimization subsequently leverages the information of the artificial intelligence step to provide suggestions for new compositions. For each cycle, 14 new syntheses were carried out and characterized. The amended dataset is subjected to a new optimization cycle.} 
    \label{fig:Fig01}
\end{figure}

Halide perovskite nanocrystals (PNCs), first demonstrated in 2014, have been rapidly improved, yielding tunability throughout the visible spectrum, quantum yields approaching 100\%, and diverse geometries and sizes.\cite{schmidt2014nontemplate} Due to their exceptional properties, PNCs have already been incorporated into diverse applications, focusing on optoelectronics such as LEDs, solar cells, and photodetectors, but also in field-effect transistors and, even more recently, photocatalysis. \cite{shamsi2019metal, abiram2022review, zhang2017high, gao2017novel, shyamal2020facets} Despite these impressive improvements, several issues impede widespread commercialization, such as stability, lead toxicity, and spectral efficiency in the blue region of the visible spectrum.\cite{schoonman2015organic, schileo2021lead, weng2022challenges} This latter effect is due to the chloride-perovskites being far from defect tolerant, resulting in extremely poor efficiencies compared to bromide- and iodide-based perovskites. \cite{ye2021defect} Another way to tune the spectral response in PNCs is through quantum confinement. Especially, 2D nanoplatelets (NPLs) are ideal in this regard, as they exhibit no inhomogeneous broadening in the confined dimension, with only incremental thickness values possible - currently between 2 and 6 monolayers (MLs). \cite{2018_Bohn} Analogous to the bulk-like Ruddlesden-Popper perovskites\cite{stoumpos2016ruddlesden}, their strong confinement can enable directional emission, boosting maximum external quantum efficiencies to 28\%.\cite{2020_Morgenstern} The quality of these colloidal quantum wells has improved significantly; however, the quantum yields are still far from unity, and reproducibility is an issue. Improving the NPL quality or that of any NC dispersion is an arduous task, involving a vast possible number of parameters relating to composition and fabrication. Synthesizing all of these is both infeasible and unnecessary, as it is possible to create robust and data-efficient predictors to describe the outcome of changes in fabrication parameters.\cite{Mayr_2022} Artificial neural networks (ANNs) have been widely used to approximate the quantities of interest,\cite{2019_Chen, 2020_Regonia} but success has also been achieved with, e.g., random forests and support vector machines. \cite{2020_Baum, 2021_Rickert} It seems clear that no algorithm is universally superior, but rather each may be more suited for specific applications.\cite{1997_Wolpert} Despite their impressive results, these methods occasionally suffer from extrapolation into areas of input space with sparse or no data. Therefore, material science is a fertile ground for applying methods with inbuilt uncertainty quantification functionalities, such as Gaussian processes (GPs), which are especially attractive given their data efficiency and robustness against overfitting.\cite{2020_Li, 2020_Zhang, 2020_Seko} These methods provide an excellent set of predictors to determine the effect that different fabrication and chemical parameters have on the resulting material properties. Nevertheless, the optimal values of these parameters are not trivially determined. This field has lately been dominated by Bayesian optimization. In this scheme, the value of the predicted objective function is weighted against the intrinsic uncertainty of the prediction to balance the exploration of new areas against the exploitation of already acquired information.\cite{2018_Balachandran, 2019_Voznyy}. However, this has required a massive experimental effort to achieve impressive results. In the future, models to optimize novel materials must focus on two aspects: incorporating multiple properties to increase the versatility of the targeted materials and reducing the number of syntheses necessary to achieve good results.

In this study, we develop a model addressing both of these issues as illustrated in \cref{fig:Fig01}. By combining Gaussian processing with a neural network and a random forest classifier, we can significantly reduce the demand for optimizing the synthesis of 2D \ce{CsPbBr3}-based NPLs. The algorithm uses three precursor amounts to predict the emission wavelength of the resulting NPLs and the quality, i.e., the homogeneity of the photoluminescence (PL) spectra. Starting from a pool of 100 initial syntheses, we carried out seven rounds of optimization. The algorithm produced 14 new precursor combinations, which were then used for synthesis and the PL of the resulting dispersions measured. For all previously synthesized NPL thicknesses (2-6 ML), we significantly reduced the width and asymmetry of the PL emission, signifying higher homogeneity. Additionally, the algorithm effectively predicted precursor combinations leading to hitherto unobtained, thicker NPLs (7 and 8 MLs). The algorithm's performance was exceptional, especially considering the small amount of necessary experimental synthesizing.\\

\section{Results and Discussion}\label{sec:results}
In contrast to typical inorganic semiconductor NC syntheses and halide perovskite quantum dots, which rely on the hot-injection method, our synthesis is based on the ligand-assisted reprecipitation (LARP) method.\cite{ShamsiReview, shamsi2017bright, 2018_Bohn} Importantly, it is conducted at room temperature in ambient atmospheres.\cite{2018_Bohn, sichert, tong2016highly} Briefly, a cesium-oleate precursor is injected into a \ce{PbBr2}-ligand (comprising oleylamine and oleic acid) solution. After approximately \SI{10}{s}, acetone, which acts as an antisolvent, is injected into the solution to induce NPL formation. After \SI{60}{s} of vigorous stirring, the reaction is terminated by centrifugation at \SI{1800}{g} for \SI{3}{min}. The supernatant is discarded, and the precipitate is redispersed in hexane. Our previous studies determined that nanoplatelets (NPLs) form when the amount of the A-site cation, in this case, cesium, is restricted.\cite{sichert} By fine-tuning the volumes of the cesium and lead-precursors and the anti-solvent, acetone, we were able to obtain nearly homogeneous dispersions of NPLs from 2 to 6 monolayers (MLs).\cite{2018_Bohn} Accordingly, we chose these three parameters as the input parameters for the learning algorithm while keeping all others, e.g., synthesis time, temperature, and humidity, essentially constant. To quantify the quality of the synthesis, we acquired and analyzed PL spectra of the resulting dispersions. We chose not to focus on the PL intensity but rather on the narrowness and the spectral position of the PL emission. Accordingly, we determined the narrowness by the usual half width at half maximum (hwhm) metric,\cite{2015_Reddy} while the symmetry was given by interpreting the spectrum as a distribution and calculating its skewness.\cite{2011_Doane}. As these two metrics have different ranges, they were normalized separately with the min-max method between 0 and 1 and subsequently added. This quantity is then normalized, yielding the figure of merit (FoM) of the respective synthesis: 

\begin{equation}\label{eq:FoM}
FoM = norm[\frac{1}{f_\mathrm{sym} + f_\mathrm{nar}}].
\end{equation}

\begin{figure}[!bt]
    \centering
    \includegraphics[width=0.9\textwidth]{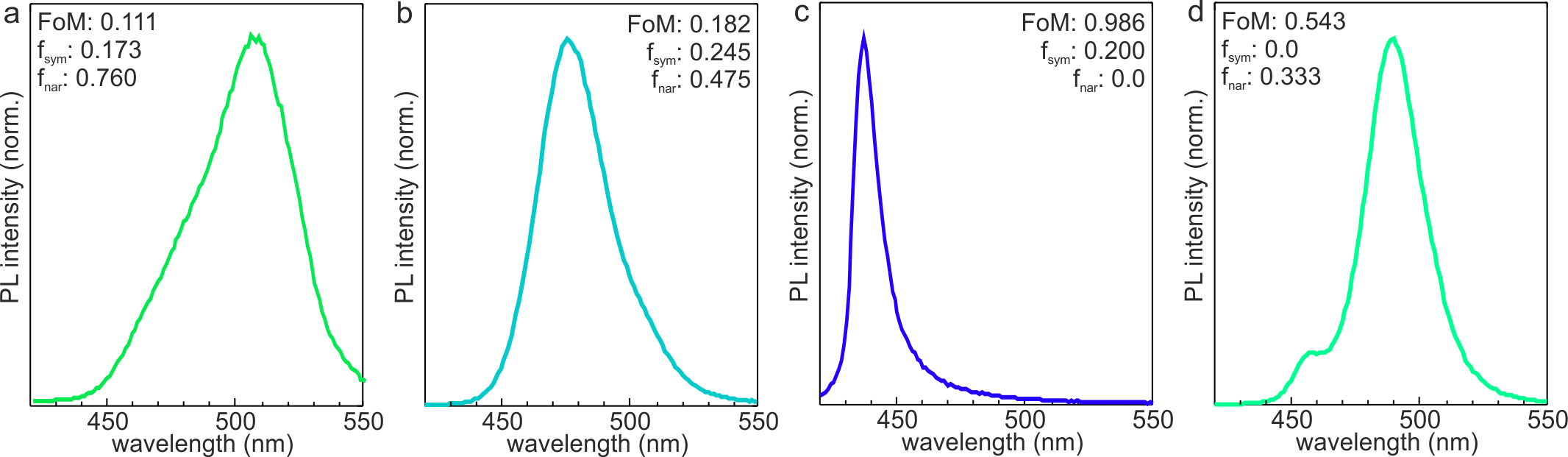}
       \caption{Examples of synthesized spectra with classification of their merit according to narrowness ($f_\mathrm{nar}$), symmetry ($f_\mathrm{sym}$) and overall quality (FoM). (a,b) Examples of poor spectra with (a) being extremely broad and (b) being quite asymmetric. (c) PL spectrum from the synthesis with the highest obtained FoM value. An extremely narrow spectrum compensates for the slight asymmetry. (d) The FoM calculation classifies this PL spectrum as very good due to a near-perfect symmetry. However, the short wavelength shoulder is a clear indicator of multiple thicknesses. This feature is predicted and preempted with the Random forest classifier.}
    \label{fig:Fig02}
\end{figure}

Accordingly, small values of $f_\mathrm{sym}$ and $f_\mathrm{sym}$ indicate good samples with a resulting FoM of 1 being a perfect spectrum and 0 a very poor one. Examples of poor spectra are shown in \cref{fig:Fig02}a,b, and the best spectrum according to the metric obtained in \cref{fig:Fig02}c. The goal of the machine-learning powered process is thus the maximization of the FoM. Several aspects must be considered when developing the full framework for guiding the NPL synthesis. Starting from a limited number of syntheses ($~100$), we have to be sure to prevent overfitting. Therefore, to predict which precursor ratios yield the highest FoM values, we adopted a Gaussian process (GP) predictor using the python package scikit-learn.\cite{2011_Pedregosa}. This choice was made because Gaussian Processes are generally more robust than others with a limited number of data. Our goal was to obtain ideal spectra for a given NPL thickness; hence, we needed a way to determine how many MLs the resulting NPLs would have. Due to the strong confinement of the NPLs, there is a robust correlation between thickness and PL emission wavelength.\cite{2018_Bohn} We implemented a neural network through the tensorflow library\cite{2015_Abadi} to predict the PL peak position based on the precursor ratios, starting from the pool of initial syntheses. We incorporated this into the overall FoM optimization pipeline as soft or hard Lagrange multipliers, constraining the PL to a given spectral window (see \Cref{tab:peaks}). In the soft case, we only incentivize and do not fully constrain PL emission near the mean of the spectral range. In contrast, in the hard case, the peak position is forced to fall on the mean, under the rationale that the suggestions will be least affected by the peak position prediction error. The hard constraints gave the best results in the initial stages of the optimization, when the peak position was still inaccurate, with the soft constraints giving better flexibility and therefore results when the peak position could be more accurately predicted. More details are given in the supporting information. 

\begin{table}[!h] 
     \caption{NPL thickness and constraints for upper and lower limits}
     \label{tab:peaks}
    
\centering
\begin{tabular}{ccccccccc}

\multicolumn{1}{c}{}                                     \\
NPL thickness [ML]  & 2   & 3   & 4   & 5   & 6   & 7   & 8        \\
\midrule
lower limit  {[}nm{]}   & 427 & 455 & 472 & 484 & 491 & 499 & 504       \\
upper limit {[}nm{]}    & 438 & 467 & 479 & 489 & 498 & 503 & 507      \\

\end{tabular}
\end{table}

In some syntheses, the PL spectrum revealed secondary peaks, which were not identified through the hwhm method, yet signify polydispersity, see \cref{fig:Fig02}d. For this synthesis, the FoM is relatively high at 0.543; however, the shoulder at \si{455 nm} is a clear sign of an inhomogeneous NPL population. To increase the data efficiency of the optimization pipeline, we trained a random forest classifier with the aid of the library scikit-learn to identify the conditions leading to polydisperse NPL distributions (see methods section for details). Importantly, while the classifier could identify roughly half of the combinations resulting in multiple peaks, it only erroneously excluded less than 5\% of the combinations yielding single peaks (see SI for more details). Consequently, viable compositions are hardly prevented from being evaluated while large areas of compositional space are cordoned off, significantly improving the algorithm's efficiency. The final set of considerations results from heuristic laboratory rules garnered from experience. Among these are minimum precursor volumes, the condition that the concentration of the Cs-precursor must be larger than the lead precursor to ensure NPL formation, and that the acetone volume is at least 30\% of the total precursor volume to ensure the function as an antisolvent.\\

These individual contributions were merged into a Bayesian Optimization Framework to deliver suggestions for promising precursor ratios for new syntheses. This approach generates suggestions focusing initially on points with high uncertainty and gradually shifting toward known areas with a high payoff. The advantage of this method is that it exploits known space and explores potentially promising points, but it completely neglects areas with a low payoff and low uncertainty. This results in a very data-efficient scheme, as only a fraction of the parameter space is explored to reveal local maxima and minima. The tradeoff between exploitation and exploration is merged into a surrogate function which is then optimized. Not many data points are initially available, so the algorithm favors exploration at the cost of exploitation. However, over time, this balance shifts with further exploitation leading to a faster optimization of the surrogate function. In this phase, already acquired data points are exploited to refine suggestions and obtain optimized peaks. In this work, we used the Expected Improvement acquisition function to define the balance between exploration and exploitation (see \Cref{sub:algorithm} for details).\\

\begin{figure}
    \includegraphics[width=0.48\textwidth]{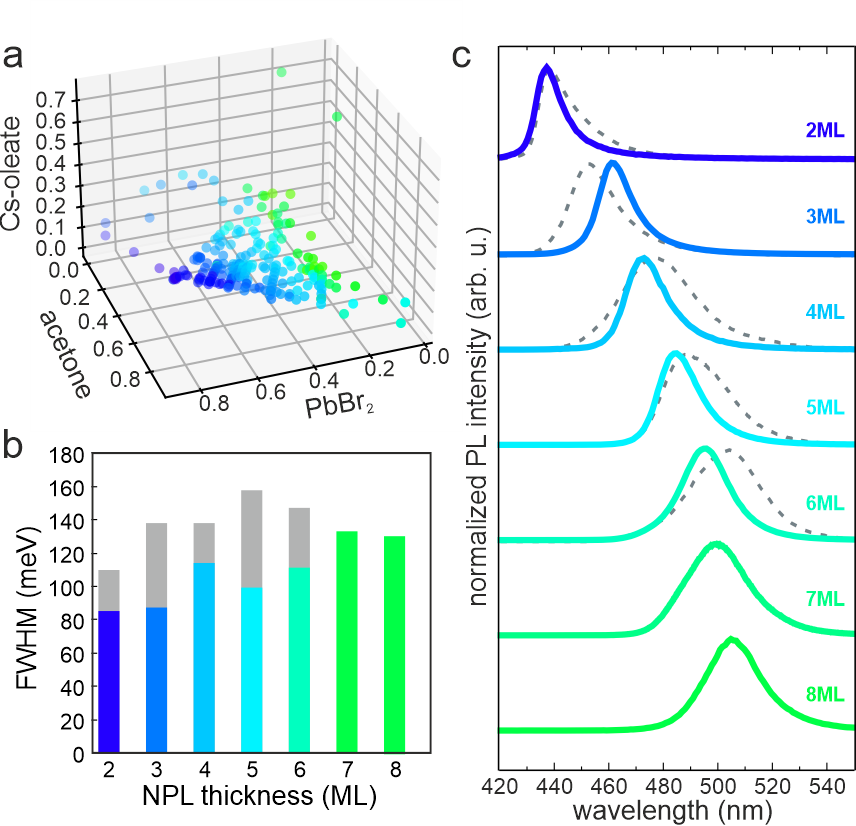}
    \caption{(a) Results of all synthesized NPL dispersions displayed in the three-dimensional precursor parameter space. The points are colored according to their respective PL emission maximum. (b) The narrowness of NPL spectra was measured by fwhm values (in meV) before optimization (gray bars) and after optimization (colored bars). A clear improvement is observed for all samples. (c) Optimized PL spectra for all NPL thicknesses (colored lines) compared to initial, typical PL spectra (dashed gray). As the 7 and 8 ML NPLs were only obtained through optimization, there are no initial spectra.}
    \label{fig:Fig03}
\end{figure}

Specifically, we generated 100.000 random three-dimensional vectors (representing the precursor ratios), adhering to the heuristic constraints as mentioned earlier. For each NPL thickness, we determined the vectors with the highest overall optimization goal not predicted to exhibit a double peak, and synthesized NPLs with these precursor settings. Overall, we ran seven cycles with 14 syntheses per cycle for a total of 220 syntheses, including the initial ones. An overview of all the syntheses is given in \cref{fig:Fig03}a,

where the syntheses are located in the three-dimensional input parameter space with the color signifying the emission wavelength. The overall trend is similar, with smaller Cs:Pb ratios leading to shorter emission wavelengths emanating from thinner NPLs. Note that the point density deviates significantly throughout the parameter space. This constitutes a visual representation of the algorithm initially exploring wide areas loosely and focusing on specific areas of high interest once it has developed a good understanding of the space. The quality of the PL spectra improved noticeably, as can be seen in SI for the case of 5 ML NPLs. The fwhm narrows considerably from 38 to 19 nm (193 to 99 meV), while the $f_\mathrm{sym}$ is reduced by 1.607 to 0, achieving perfect symmetry. Additionally, the spectral position of the PL maximum shifts gradually from 490 nm to 486 nm. In terms of the FoM, this corresponds to an improvement from 0.429 to 0.916. Similar improvement was achieved for all previously established NPL thicknesses, as shown in \cref{fig:Fig03}b, where the fwhm before (gray bars) and after optimization (colored) are shown for all thicknesses. The optimized spectra are displayed as colored curves in \cref{fig:Fig03}c, where the dashed gray ones are from typical initial syntheses. All optimized spectra are significantly narrower and less asymmetric. All peak maxima exhibited a blueshift with the sole exception of the 3 ML NPLs. The improvement can also be seen in the FoM values, which increased substantially for all thicknesses. These are displayed in  SI along with the precursor amounts (both new and old). While there are only subtle shifts in the precursor amounts for some thicknesses, there are also significant deviations, such as for the 4 ML and 6 ML NPLs. This demonstrates the advantage of the employed approach, as it is unlikely these precursor ratios would have been heuristically selected. The quality of the spectra is noticeable even when compared to the best spectra obtained from the initial NPL synthesis (see SI).\cite{2018_Bohn}\\

\begin{figure}[t]
    \centering
    \includegraphics[width=0.95\textwidth]{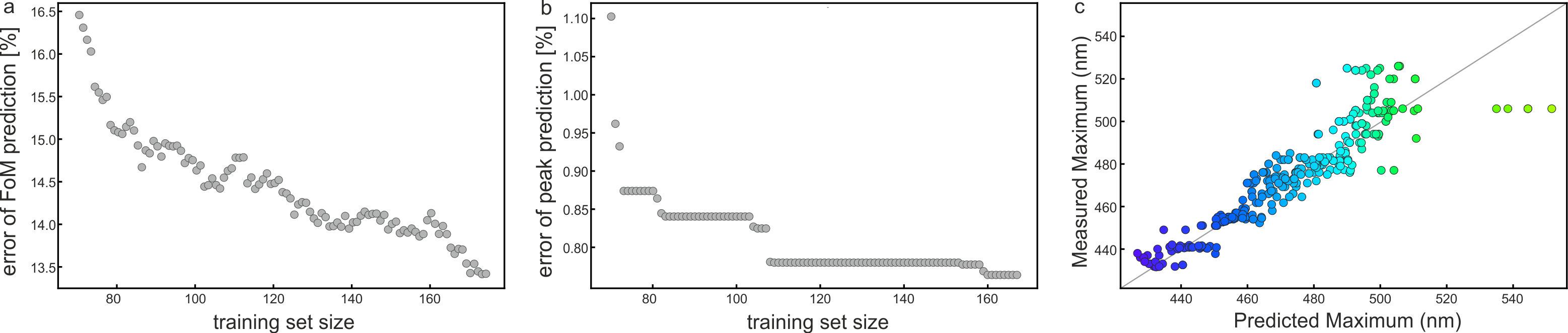}
    \caption{(a) Evolution of error for the FoM prediction with an increasing number of training samples. The error would likely decrease even further with additional syntheses. (b) Evolution of error for the peak position predictor with an increasing number of training samples. The initially low error has almost reached a plateau, confirming the high efficiency of the algorithm in predicting the resulting PL emission maxima. (c) The final accuracy of the peak position predictor. The outlier points correspond to rare occurrences in the dataset and hardly decrease the overall evaluation. All results were generated with a 90-10 training validation split and 10-fold validation.}
    \label{fig:Fig04}
\end{figure}

The algorithm performed exceptionally well, as shown in \cref{fig:Fig04}. With every new synthesis added to the training sets, the error in predicting the quality factor decreased substantially (\cref{fig:Fig04}a). As the downward trend is still steep at the end of the implemented training data, the algorithm would probably improve further with more syntheses. In contrast, the predictor for the resulting PL emission wavelength improved rapidly for the first 30-40 syntheses, continued by a slower improvement (\cref{fig:Fig04}b). This shows how well the precursor amounts define the PL emission of the resulting NPLs and that the algorithm could adapt quickly. The difference between predicted and measured PL emission maxima is highlighted in \cref{fig:Fig04}c. Here, there is a very strong grouping of the points along the diagonal (indicated as grey line), indicative of very high accuracy. Interestingly, the algorithm predicted several precursor combinations to result in spectra with emission maxima at wavelengths longer than \SI{494}{nm}, which corresponds to 6 ML NPLs. While not all of these resulted in NPLs, we were able to obtain two separate dispersions with distinct emission maxima at \SI{501}{nm} and \SI{505}{nm}, which we could assign to 7 ML and 8 ML NPLs, respectively and confirm through TEM imaging (see SI). We must note that the 8 ML NPL dispersion also contains small amounts of slightly thicker NPLs.

Nevertheless, the algorithm enabled us to obtain specific NPL thicknesses for the first time, which also helped determine important excitonic properties of the strongly confined asymmetric nanocrystals.\cite{2022_Gramlich}. This is very impressive considering a) the significantly low amount of data (\textit{i.e.} syntheses) being fed into the algorithm and b) the fact that no data points in this spectral region were provided, proving the strong extrapolative abilities of our pipeline. We carried out only 220 syntheses, which is more than an order of magnitude less than in similar studies.\cite{2019_Voznyy} Our previous experience with the synthesis certainly helped streamline the process. Accordingly, in the next step, we will apply the algorithm to a novel synthesis, which we have not yet had experience with, and test how well it performs in this case and which parts might need further optimization. Also, to further increase turnaround time and enhance reproducibility, we will look into merging the algorithm with high throughput, automated synthesis via microfluidics with temperature control.\cite{2019_Sun, 2020_Lignos}
\WFclear 

\section{Conclusion}

In summary, we have implemented a novel machine-learning method to obtain \ce{CsPbBr3} NPLs of a given thickness and with high optical quality with a vastly reduced amount of data. The algorithm, comprising three machine-learning models merged with Bayesian optimization, can optimize the spectra' quality, thus the monodispersity of the samples and the desired emission wavelength, \textit{i.e.} the thickness of the resulting NPLs. Additionally, predicting and taking into account the purity of the resulting spectrum and incorporating additional heuristic constraints helped decrease the data pressure, meaning that the algorithm was able to rapidly improve the quality of NPL dispersions while requiring only at most three optimization cycles per NLP. In total, only 220 samples were required to obtain these results. Moreover, the model was able to identify precursor ratios leading to hitherto unobtainable NPL thicknesses, proving that the models could learn correlations beyond the training set bounds. The model can easily be expanded to account for additional synthetic parameters, such as temperature and humidity, and optimize quantum yields or stability. The next steps will constitute testing the versatility of the algorithm by transferring it to an unknown synthesis and implementing a high throughput automated synthesis via microfluidics to speed up the process and ensure higher reliability.

\section{Experimental Section}

\subsection{Synthesis}
Materials: \ce{Cs2CO3} (cesium carbonate, 99\%), \ce{PbBr2} (lead(II)bromide, $>98\%$), oleic acid (technical grad 90\%), oleylamine (technical grade 70\%), acetone (for HPLC, $>99.9\%$), toluene (for HPLC, $>99.9\%$) and hexane (for HPLC, $>97.0\%$, GC) were purchased from Sigma Aldrich and used without further purification.\\
\hfill

The \ce{PbBr2} ligand solution and Cs-oleate were prepared according to Bohn et al.\cite{2018_Bohn}
The synthesis procedure remained the same for all used synthesis parameters. It is presented in the following:\\
The synthesis was carried out under an ambient atmosphere at room temperature. The synthesis ratios were multiplied by 3 to get reasonable volumes to work with in the laboratory. A reaction glass was loaded with \ce{PbBr2}-precursor solution and Cs-oleate was immediately added under vigorous stirring. After 10 s, acetone was added quickly, and the reaction mixture was stirred for 1 min. Afterward, the mixture was centrifuged at 4000\,rpm for 3\,min, and the precipitate was redispersed in hexane (2\,ml). Immediately after synthesis, the samples were optically characterized with a commercial spectrometer (FLOUROMAX-Plus, HORIBA). 

\subsection{Algorithm details}\label{sub:algorithm}
For the Gaussian Processes predicting the FoM the kernel employed was the sum of a radial basis function (RBF) and a Matern covariance function whose parameters were optimized by maximizing the log-marginal-likelihood (LML).\\
\hfill

The neural network predicting the peak position consisted of three hidden layers with five neurons each. The first was activated using the Exponential Linear Unit (ELU) and the subsequent two with the Rectified Linear Unit (ReLU). The output layer activation was ELU. The weight initialization was
the He uniform variance scaling initializer, and the optimizer used in this case was Root Mean Squared Propagation (RMSDrop) with an exponentially decaying learning rate.\\
\hfill

The random forest classifier was trained with the constraint that every leaf should contain at least ten samples, and the optimization criterion reflecting the purity of each leaf is the gini index. \\
\hfill

The Bayesian optimization scheme used the expected improvement acquisition function, constrained suitably with the constraints described in \cref{sec:results}.

\medskip
\textbf{Supporting Information} \par 
Supporting Information is available from the Wiley Online Library or from the author.

\medskip
\textbf{Acknowledgements} \par 
The writers acknowledge funding from the following sources: 

\vspace{5mm}

This project was funded by: the European Research Council Horizon 2020 through the ERC Grant Agreement PINNACLE (759744), by the Deutsche Forschungsgemeinsschaft (DFG) under Germany’s Excellence Strategy EXC 2089/1-390776260, by the Bavarian State Ministry of Science, Research and Arts through the grant “Solar Technologies go Hybrid (SolTech)”,

the Project ProperPhotoMile, supported under the umbrella of SOLAR-ERA.NET Cofund 2 by The Spanish Ministry of Science and Education and the AEI under the project PCI2020-112185 and CDTI project number IDI-20210171; the Federal Ministry for Economic Affairs and Energy on the basis of a decision by the German Bundestag project number FKZ 03EE1070B and FKZ 03EE1070A and the Israel Ministry of Energy with project number 220-11-031.  SOLAR-ERA.NET is supported by the European Commission within the EU Framework Programme for Research and Innovation HORIZON 2020 (Cofund ERA-NET Action, N 786483),

and by the “EXC 2089: e-conversion” DFG-cluster of excellence [project number: 390776260, https://www.e-conversion.de/]. "ARTEMIS" - TUM innovation network, Technical University of Munich funded through the German Excellence Initiative and the state of Bavaria.

 \vspace{5mm}


\bibliographystyle{unsrtnat}


\begin{thebibliography}{35}
\providecommand{\natexlab}[1]{#1}
\providecommand{\url}[1]{\texttt{#1}}
\expandafter\ifx\csname urlstyle\endcsname\relax
  \providecommand{\doi}[1]{doi: #1}\else
  \providecommand{\doi}{doi: \begingroup \urlstyle{rm}\Url}\fi

\bibitem[Schmidt et~al.(2014)Schmidt, Perteg{\'a}s, Gonz{\'a}lez-Carrero,
  Malinkiewicz, Agouram, Minguez~Espallargas, Bolink, Galian, and
  P{\'e}rez-Prieto]{schmidt2014nontemplate}
Luciana~C Schmidt, Antonio Perteg{\'a}s, Soranyel Gonz{\'a}lez-Carrero, Olga
  Malinkiewicz, Said Agouram, Guillermo Minguez~Espallargas, Henk~J Bolink,
  Raquel~E Galian, and Julia P{\'e}rez-Prieto.
\newblock Nontemplate synthesis of \ce{CH3NH3PbBr3} perovskite nanoparticles.
\newblock \emph{Journal of the American Chemical Society}, 136\penalty0
  (3):\penalty0 850--853, 2014.

\bibitem[Shamsi et~al.(2019{\natexlab{a}})Shamsi, Urban, Imran, De~Trizio, and
  Manna]{shamsi2019metal}
Javad Shamsi, Alexander~S Urban, Muhammad Imran, Luca De~Trizio, and Liberato
  Manna.
\newblock Metal halide perovskite nanocrystals: synthesis, post-synthesis
  modifications, and their optical properties.
\newblock \emph{Chemical reviews}, 119\penalty0 (5):\penalty0 3296--3348,
  2019{\natexlab{a}}.

\bibitem[Abiram et~al.(2022)Abiram, Thanihaichelvan, Ravirajan, and
  Velauthapillai]{abiram2022review}
Gnanasampanthan Abiram, Murugathas Thanihaichelvan, Punniamoorthy Ravirajan,
  and Dhayalan Velauthapillai.
\newblock Review on perovskite semiconductor field--effect transistors and
  their applications.
\newblock \emph{Nanomaterials}, 12\penalty0 (14):\penalty0 2396, 2022.

\bibitem[Zhang et~al.(2017)Zhang, Wang, Zhang, Jiang, Gao, Jin, and
  Liu]{zhang2017high}
Jingru Zhang, Qian Wang, Xisheng Zhang, Jiexuan Jiang, Zhenfei Gao, Zhiwen Jin,
  and Shengzhong~Frank Liu.
\newblock High-performance transparent ultraviolet photodetectors based on
  inorganic perovskite \ce{CsPbCl3} nanocrystals.
\newblock \emph{RSC advances}, 7\penalty0 (58):\penalty0 36722--36727, 2017.

\bibitem[Gao et~al.(2017)Gao, Xi, Zhou, Zhao, Wu, Wang, Guo, and
  Xu]{gao2017novel}
Ge~Gao, Qiaoyue Xi, Hua Zhou, Yongxia Zhao, Cunqi Wu, Lidan Wang, Pengran Guo,
  and Jingwei Xu.
\newblock Novel inorganic perovskite quantum dots for photocatalysis.
\newblock \emph{Nanoscale}, 9\penalty0 (33):\penalty0 12032--12038, 2017.

\bibitem[Shyamal et~al.(2020)Shyamal, Dutta, Das, Sen, Chakraborty, and
  Pradhan]{shyamal2020facets}
Sanjib Shyamal, Sumit~Kumar Dutta, Tisita Das, Suvodeep Sen, Sudip Chakraborty,
  and Narayan Pradhan.
\newblock Facets and defects in perovskite nanocrystals for photocatalytic
  \ce{CO2} reduction.
\newblock \emph{The journal of physical chemistry letters}, 11\penalty0
  (9):\penalty0 3608--3614, 2020.

\bibitem[Schoonman(2015)]{schoonman2015organic}
J~Schoonman.
\newblock Organic--inorganic lead halide perovskite solar cell materials: a
  possible stability problem.
\newblock \emph{Chemical Physics Letters}, 619:\penalty0 193--195, 2015.

\bibitem[Schileo and Grancini(2021)]{schileo2021lead}
Giorgio Schileo and Giulia Grancini.
\newblock Lead or no lead? availability, toxicity, sustainability and
  environmental impact of lead-free perovskite solar cells.
\newblock \emph{Journal of Materials Chemistry C}, 9\penalty0 (1):\penalty0
  67--76, 2021.

\bibitem[Weng et~al.(2022)Weng, Yu, Zhou, Lin, Han, Wang, Huang, Liu, Hu, Liu,
  et~al.]{weng2022challenges}
Shuchen Weng, Guicheng Yu, Chao Zhou, Fang Lin, Yonglei Han, Hao Wang, Xiaoxi
  Huang, Xiaoyuan Liu, Hanlin Hu, Wei Liu, et~al.
\newblock Challenges and opportunities for the blue perovskite quantum dot
  light-emitting diodes.
\newblock \emph{Crystals}, 12\penalty0 (7):\penalty0 929, 2022.

\bibitem[Ye et~al.(2021)Ye, Byranvand, Mart{\'\i}nez, Hoye, Saliba, and
  Polavarapu]{ye2021defect}
Junzhi Ye, Mahdi~Malekshahi Byranvand, Clara~Otero Mart{\'\i}nez, Robert~LZ
  Hoye, Michael Saliba, and Lakshminarayana Polavarapu.
\newblock Defect passivation in lead-halide perovskite nanocrystals and thin
  films: toward efficient leds and solar cells.
\newblock \emph{Angewandte Chemie}, 133\penalty0 (40):\penalty0 21804--21828,
  2021.

\bibitem[Bohn et~al.(2018)Bohn, Tong, Gramlich, Lai, Doblinger, Wang, Hoye,
  Muller~Buschbaum, Stranks, Urban, Polavarapu, and Feldmann]{2018_Bohn}
Bernhard~J. Bohn, Yu~Tong, Moritz Gramlich, May~Ling Lai, Markus Doblinger, Kun
  Wang, Robert L.~Z. Hoye, Peter Muller~Buschbaum, Samuel~D. Stranks,
  Alexander~S. Urban, Lakshminarayana Polavarapu, and Jochen Feldmann.
\newblock Boosting tunable blue luminescence of halide perovskite nanoplatelets
  through postsynthetic surface trap repair.
\newblock \emph{Nano Letters}, 18\penalty0 (8):\penalty0 5231--5238, 2018.
\newblock \doi{10.1021/acs.nanolett.8b02190}.
\newblock URL \url{https://doi.org/10.1021/acs.nanolett.8b02190}.
\newblock PMID: 29990435.

\bibitem[Stoumpos et~al.(2016)Stoumpos, Cao, Clark, Young, Rondinelli, Jang,
  Hupp, and Kanatzidis]{stoumpos2016ruddlesden}
Constantinos~C Stoumpos, Duyen~H Cao, Daniel~J Clark, Joshua Young, James~M
  Rondinelli, Joon~I Jang, Joseph~T Hupp, and Mercouri~G Kanatzidis.
\newblock Ruddlesden--popper hybrid lead iodide perovskite 2d homologous
  semiconductors.
\newblock \emph{Chemistry of Materials}, 28\penalty0 (8):\penalty0 2852--2867,
  2016.

\bibitem[Morgenstern et~al.(2020)Morgenstern, Lampe, Naujoks, Jurow, Liu,
  Urban, and Brütting]{2020_Morgenstern}
Thomas Morgenstern, Carola Lampe, Tassilo Naujoks, Matthew Jurow, Yi~Liu,
  Alexander~S. Urban, and Wolfgang Brütting.
\newblock Elucidating the performance limits of perovskite nanocrystal light
  emitting diodes.
\newblock \emph{Journal of Luminescence}, 220:\penalty0 116939, 2020.
\newblock ISSN 0022-2313.
\newblock \doi{https://doi.org/10.1016/j.jlumin.2019.116939}.
\newblock URL
  \url{https://www.sciencedirect.com/science/article/pii/S0022231319313742}.

\bibitem[Mayr et~al.(2022)Mayr, Harth, Kouroudis, Rinderle, and
  Gagliardi]{Mayr_2022}
Felix Mayr, Milan Harth, Ioannis Kouroudis, Michael Rinderle, and Alessio
  Gagliardi.
\newblock Machine learning and optoelectronic materials discovery: A growing
  synergy.
\newblock \emph{The Journal of Physical Chemistry Letters}, 13:\penalty0
  1940--1951, 2022.

\bibitem[Chen and Pao(2019)]{2019_Chen}
Hsin-An Chen and Chun-Wei Pao.
\newblock Fast and accurate artificial neural network potential model for
  mapbi3 perovskite materials.
\newblock \emph{ACS omega}, 4\penalty0 (6):\penalty0 10950--10959, 2019.

\bibitem[Regonia et~al.(2020)Regonia, Pelicano, Tani, Ishizumi, Yanagi, and
  Ikeda]{2020_Regonia}
Paul~Rossener Regonia, Christian~Mark Pelicano, Ryosuke Tani, Atsushi Ishizumi,
  Hisao Yanagi, and Kazushi Ikeda.
\newblock Predicting the band gap of zno quantum dots via supervised machine
  learning models.
\newblock \emph{Optik}, 207:\penalty0 164469, 2020.

\bibitem[Baum et~al.(2020)Baum, Pretto, Koche, and Santos]{2020_Baum}
Fabio Baum, Tatiane Pretto, Ariadne Koche, and Marcos Jose~Leite Santos.
\newblock Machine learning tools to predict hot injection syntheses outcomes
  for ii--vi and iv--vi quantum dots.
\newblock \emph{The Journal of Physical Chemistry C}, 124\penalty0
  (44):\penalty0 24298--24305, 2020.

\bibitem[Rickert et~al.(2021)Rickert, Hayta, Selle, Kouroudis, Harth,
  Gagliardi, and Lieleg]{2021_Rickert}
Carolin~A Rickert, Elif~N Hayta, Daniel~M Selle, Ioannis Kouroudis, Milan
  Harth, Alessio Gagliardi, and Oliver Lieleg.
\newblock Machine learning approach to analyze the surface properties of
  biological materials.
\newblock \emph{ACS Biomaterials Science \& Engineering}, 7\penalty0
  (9):\penalty0 4614--4625, 2021.

\bibitem[Wolpert and Macready(1997)]{1997_Wolpert}
David~H Wolpert and William~G Macready.
\newblock No free lunch theorems for optimization.
\newblock \emph{IEEE transactions on evolutionary computation}, 1\penalty0
  (1):\penalty0 67--82, 1997.

\bibitem[Li et~al.(2020)Li, Achenie, and Xin]{2020_Li}
Zheng Li, Luke~EK Achenie, and Hongliang Xin.
\newblock An adaptive machine learning strategy for accelerating discovery of
  perovskite electrocatalysts.
\newblock \emph{ACS Catalysis}, 10\penalty0 (7):\penalty0 4377--4384, 2020.

\bibitem[Zhang and Xu(2020)]{2020_Zhang}
Yun Zhang and Xiaojie Xu.
\newblock Machine learning lattice constants for cubic perovskite compounds.
\newblock \emph{ChemistrySelect}, 5\penalty0 (32):\penalty0 9999--10009, 2020.

\bibitem[Seko and Ishiwata(2020)]{2020_Seko}
Atsuto Seko and Shintaro Ishiwata.
\newblock Prediction of perovskite-related structures in $\ce{ACuO}_{3-x}$
  (a=ca, sr, ba, sc, y, la) using density functional theory and bayesian
  optimization.
\newblock \emph{Physical Review B}, 101\penalty0 (13):\penalty0 134101, 2020.

\bibitem[Balachandran et~al.(2018)Balachandran, Kowalski, Sehirlioglu, and
  Lookman]{2018_Balachandran}
Prasanna~V Balachandran, Benjamin Kowalski, Alp Sehirlioglu, and Turab Lookman.
\newblock Experimental search for high-temperature ferroelectric perovskites
  guided by two-step machine learning.
\newblock \emph{Nature communications}, 9\penalty0 (1):\penalty0 1--9, 2018.

\bibitem[Voznyy et~al.(2019)Voznyy, Levina, Fan, Askerka, Jain, Choi,
  Ouellette, Todorovic, Sagar, and Sargent]{2019_Voznyy}
Oleksandr Voznyy, Larissa Levina, James~Z Fan, Mikhail Askerka, Ankit Jain,
  Min-Jae Choi, Olivier Ouellette, Petar Todorovic, Laxmi~K Sagar, and Edward~H
  Sargent.
\newblock Machine learning accelerates discovery of optimal colloidal quantum
  dot synthesis.
\newblock \emph{ACS nano}, 13\penalty0 (10):\penalty0 11122--11128, 2019.

\bibitem[Shamsi et~al.(2019{\natexlab{b}})Shamsi, Urban, Imran, De~Trizio, and
  Manna]{ShamsiReview}
Javad Shamsi, Alexander~S. Urban, Muhammad Imran, Luca De~Trizio, and Liberato
  Manna.
\newblock Metal halide perovskite nanocrystals: Synthesis, post-synthesis
  modifications, and their optical properties.
\newblock \emph{Chemical Reviews}, 119\penalty0 (5):\penalty0 3296--3348,
  2019{\natexlab{b}}.
\newblock \doi{10.1021/acs.chemrev.8b00644}.
\newblock URL \url{https://doi.org/10.1021/acs.chemrev.8b00644}.
\newblock PMID: 30758194.

\bibitem[Shamsi et~al.(2017)Shamsi, Rastogi, Caligiuri, Abdelhady, Spirito,
  Manna, and Krahne]{shamsi2017bright}
Javad Shamsi, Prachi Rastogi, Vincenzo Caligiuri, Ahmed~L Abdelhady, Davide
  Spirito, Liberato Manna, and Roman Krahne.
\newblock Bright-emitting perovskite films by large-scale synthesis and
  photoinduced solid-state transformation of \ce{CsPbBr3} nanoplatelets.
\newblock \emph{ACS Nano}, 11\penalty0 (10):\penalty0 10206--10213, 2017.
\newblock \doi{10.1021/acsnano.7b04761}.

\bibitem[Sichert et~al.(2015)Sichert, Tong, Mutz, Vollmer, Fischer, Milowska,
  Garc{\'\i}a~Cortadella, Nickel, Cardenas-Daw, Stolarczyk, et~al.]{sichert}
Jasmina~A Sichert, Yu~Tong, Niklas Mutz, Mathias Vollmer, Stefan Fischer,
  Karolina~Z Milowska, Ramon Garc{\'\i}a~Cortadella, Bert Nickel, Carlos
  Cardenas-Daw, Jacek~K Stolarczyk, et~al.
\newblock Quantum size effect in organometal halide perovskite nanoplatelets.
\newblock \emph{Nano letters}, 15\penalty0 (10):\penalty0 6521--6527, 2015.

\bibitem[Tong et~al.(2016)Tong, Bladt, Ayg{\"u}ler, Manzi, Milowska,
  Hintermayr, Docampo, Bals, Urban, Polavarapu, et~al.]{tong2016highly}
Yu~Tong, Eva Bladt, Meltem~F Ayg{\"u}ler, Aurora Manzi, Karolina~Z Milowska,
  Verena~A Hintermayr, Pablo Docampo, Sara Bals, Alexander~S Urban,
  Lakshminarayana Polavarapu, et~al.
\newblock Highly luminescent cesium lead halide perovskite nanocrystals with
  tunable composition and thickness by ultrasonication.
\newblock \emph{Angewandte Chemie International Edition}, 55\penalty0
  (44):\penalty0 13887--13892, 2016.

\bibitem[Kumar~Reddy and Sagar(2015)]{2015_Reddy}
Andra~Naresh Kumar~Reddy and Dasari~Karuna Sagar.
\newblock Half-width at half-maximum, full-width at half-maximum analysis for
  resolution of asymmetrically apodized optical systems with slit apertures.
\newblock \emph{Pramana}, 84\penalty0 (1):\penalty0 117--126, 2015.

\bibitem[Doane and Seward(2011)]{2011_Doane}
David~P Doane and Lori~E Seward.
\newblock Measuring skewness: a forgotten statistic?
\newblock \emph{Journal of statistics education}, 19\penalty0 (2), 2011.

\bibitem[Pedregosa et~al.(2011)Pedregosa, Varoquaux, Gramfort, Michel, Thirion,
  Grisel, Blondel, Prettenhofer, Weiss, Dubourg, Vanderplas, Passos,
  Cournapeau, Brucher, Perrot, and Duchesnay]{2011_Pedregosa}
F.~Pedregosa, G.~Varoquaux, A.~Gramfort, V.~Michel, B.~Thirion, O.~Grisel,
  M.~Blondel, P.~Prettenhofer, R.~Weiss, V.~Dubourg, J.~Vanderplas, A.~Passos,
  D.~Cournapeau, M.~Brucher, M.~Perrot, and E.~Duchesnay.
\newblock Scikit-learn: Machine learning in {P}ython.
\newblock \emph{Journal of Machine Learning Research}, 12:\penalty0 2825--2830,
  2011.

\bibitem[Abadi et~al.(2015)Abadi, Agarwal, Barham, Brevdo, Chen, Citro,
  Corrado, Davis, Dean, Devin, Ghemawat, Goodfellow, Harp, Irving, Isard, Jia,
  Jozefowicz, Kaiser, Kudlur, Levenberg, Man\'{e}, Monga, Moore, Murray, Olah,
  Schuster, Shlens, Steiner, Sutskever, Talwar, Tucker, Vanhoucke, Vasudevan,
  Vi\'{e}gas, Vinyals, Warden, Wattenberg, Wicke, Yu, and Zheng]{2015_Abadi}
Mart\'{i}n Abadi, Ashish Agarwal, Paul Barham, Eugene Brevdo, Zhifeng Chen,
  Craig Citro, Greg~S. Corrado, Andy Davis, Jeffrey Dean, Matthieu Devin,
  Sanjay Ghemawat, Ian Goodfellow, Andrew Harp, Geoffrey Irving, Michael Isard,
  Yangqing Jia, Rafal Jozefowicz, Lukasz Kaiser, Manjunath Kudlur, Josh
  Levenberg, Dandelion Man\'{e}, Rajat Monga, Sherry Moore, Derek Murray, Chris
  Olah, Mike Schuster, Jonathon Shlens, Benoit Steiner, Ilya Sutskever, Kunal
  Talwar, Paul Tucker, Vincent Vanhoucke, Vijay Vasudevan, Fernanda Vi\'{e}gas,
  Oriol Vinyals, Pete Warden, Martin Wattenberg, Martin Wicke, Yuan Yu, and
  Xiaoqiang Zheng.
\newblock {TensorFlow}: Large-scale machine learning on heterogeneous systems,
  2015.
\newblock URL \url{https://www.tensorflow.org/}.
\newblock Software available from tensorflow.org.

\bibitem[Gramlich et~al.(2022)Gramlich, Swift, Lampe, Lyons, Döblinger, Efros,
  Sercel, and Urban]{2022_Gramlich}
Moritz Gramlich, Michael~W. Swift, Carola Lampe, John~L. Lyons, Markus
  Döblinger, Alexander~L. Efros, Peter~C. Sercel, and Alexander~S. Urban.
\newblock Dark and bright excitons in halide perovskite nanoplatelets.
\newblock \emph{Advanced Science}, 9\penalty0 (5):\penalty0 2103013, 2022.
\newblock \doi{https://doi.org/10.1002/advs.202103013}.
\newblock URL
  \url{https://onlinelibrary.wiley.com/doi/abs/10.1002/advs.202103013}.

\bibitem[Sun et~al.(2019)Sun, Hartono, Ren, Oviedo, Buscemi, Layurova, Chen,
  Ogunfunmi, Thapa, Ramasamy, et~al.]{2019_Sun}
Shijing Sun, Noor~TP Hartono, Zekun~D Ren, Felipe Oviedo, Antonio~M Buscemi,
  Mariya Layurova, De~Xin Chen, Tofunmi Ogunfunmi, Janak Thapa, Savitha
  Ramasamy, et~al.
\newblock Accelerated development of perovskite-inspired materials via
  high-throughput synthesis and machine-learning diagnosis.
\newblock \emph{Joule}, 3\penalty0 (6):\penalty0 1437--1451, 2019.

\bibitem[Lignos et~al.(2020)Lignos, Maceiczyk, Kovalenko, and
  Stavrakis]{2020_Lignos}
Ioannis Lignos, Richard~M. Maceiczyk, Maksym~V. Kovalenko, and Stavros
  Stavrakis.
\newblock Tracking the fluorescence lifetimes of cesium lead halide perovskite
  nanocrystals during their synthesis using a fully automated optofluidic
  platform.
\newblock \emph{Chemistry of Materials}, 32\penalty0 (1):\penalty0 27--37,
  2020.
\newblock \doi{10.1021/acs.chemmater.9b03438}.

\end{thebibliography}

\end{document}